# Qubit-Wise Architecture Search Method for Variational Quantum Circuits

Jialin Chen[1], Zhiqiang Cai[2], Ke Xu[1], Di Wu[1], Wei Cao[1]
[1] *Institute of Big Data, Fudan University, P. R. China*
[2] *School of Microelectronics, Fudan University, P. R. China*

## Abstract

Considering the noise level limit, one crucial aspect for quantum machine learning is to design a high-performing variational quantum circuit architecture with small number of quantum gates. As the classical neural architecture search (NAS), quantum architecture search methods (QAS) employ methods like reinforcement learning, evolutionary algorithms and supernet optimization to improve the search efficiency. In this paper, we propose a novel qubit-wise architecture search (QWAS) method, which progressively search one-qubit configuration per stage, and combine with Monte Carlo Tree Search algorithm to find good quantum architectures by partitioning the search space into several good and bad subregions. The numerical experimental results indicate that our proposed method can balance the exploration and exploitation of circuit performance and size in some real-world tasks, such as MNIST, Fashion and MOSI. As far as we know, QWAS achieves the state-of-art results of all tasks in the terms of accuracy and circuit size.

## 1. Introduction

In the current era of noisy intermediate-scale quantum (NISQ) [Preskill, 2018], the advent of variational quantum algorithm (VQA) has been proved the most important research area for quantum machine learning. A key component of these algorithms is the use of variational quantum circuits (VQCs) [Mitarai *et al.*, 2018]. VQCs allow for the tuning of quantum circuit parameters through classical optimization methods. They have been validated experimentally for various small-scale learning problems, demonstrating potential advantages in domains ranging from chemistry to machine learning tasks [Kandala *et al.*, 2017; Cao *et al.*, 2019; Cong *et al.*, 2019; Henderson *et al.*, 2020].

However, considering the noise limit of NISQ hardware, one crucial aspect for quantum machine learning is to design a high-performing and robust variational quantum circuit architecture [Holmes *et al.*, 2022]. Many established ansatze typically comprise repeated layers with a fixed topology of parameterized and non-parameterized gate [Schuld *et al.*, 2020]. To develop a strategy to design VQC in an automated way, i.e. quantum architecture search (QAS), some researchers have turned their attention to the classical Neural Architecture Search (NAS) framework. NAS focuses on automating the design of neural network structures [Elsken *et al.*, 2019], but often grapple with the challenge of evaluating a vast number of possible network architectures. The Monte Carlo Tree Search (MCTS) algorithm addresses this issue by iteratively exploring and evaluating segments of the search space, thereby identifying promising neural network structures without exhaustive enumeration [Silver *et al.*, 2016; Wang *et al.*, 2020]. However, the efficiency of the search is significantly influenced by the manually predefined action space before the tree construction. To address this issue, [Wang *et al.*, 2021] proposed an improved MCTS-based algorithm called Latent Action Neural Architecture Search (LaNAS) that learns a latent action space that best fits the problem to be solved.

To bring the gap between NAS and QAS, [Zhang *et al.*, 2020] applied the NAS to the domain of quantum circuit design. [Pirhooshyaran *et al.*, 2021] proposed several search strategies including random search, reinforcement learning and Bayesian optimization, but all of them cannot change the number of parameterized quantum gates which limiting the ability to find more compact architectures. [Ostaszewski *et al.*, 2021]propose a reinforcement learning algorithm that autonomously explores the space of possible ansatzes, that an agent starts from an empty circuit and add a quantum gate per step as actions. However, this method depends on the quality of the reward function and only valid for VQE problems. [Wang *et al.*, 2023] present a nested MCTS method to iteratively selects the best subspace at each layer and constructs a new Monte Carlo tree from the selected tree node. However, rebuilding a tree at each layer incurs additional computational overhead and makes it impossible to share parameters throughout the entire search.

In addition, to reduce the computational cost of QAS, a hopeful direction is the application of the supernet methodology in the quantum area [Du *et al.*, 2022]. A supernet [Liu *et al.*, 2018] is a conceptual framework that supports the simultaneous training of multiple network architectures, thus providing a more efficient strategy to traverse the broad search space characteristic of NAS. [Wang *et al.*, 2022a] pro-pose to train a large SuperCircuit, and its sub-circuit performance is estimated with parameters inherited from SuperCircuit. Obviously, the topology and thus performance of sub-circuit is restricted to that of SuperCircuit. A primary concern is the necessity for fine-tuning the supernet architecture to align effectively with specific tasks.

In response to these challenges, we propose a novel QWAS framework for the qubit-wise search method of VQC design. Since VQCs are constructed based on quantum gates, they can be seen as pictures if the state of each gate involved represents a pixel. Then, imagine the function of QWAS is something like to make a random picture meaningful by changing a little pixel row by row. Formally, QWAS progressively optimize the part of circuit on a selected qubit per stage, then use MCTS to partition the search space for next stage to speed up the search process. Moreover, by introducing noise-adaptive terms, our method can produce more trainability and robust quantum circuit designs suitable for NISQ devices.

Our main contributions are:
- We devise a novel qubit-wise strategy to progressively search and optimize one-qubit configuration of the BaseNet, leading to the large improvement of the performance.
- We introduce noise-adaptive terms into Monte Carlo Tree Search algorithm to find more efficient and trainable quantum architectures
- We validate by extensive experiments that QWAS is able to balance the exploration and exploitation of circuit performance and size. In the context of QML, QWAS achieves the state-of-art results of all tasks.

## 2. Background

### 2.1 Variational Quantum Circuits (VQC)

The operation of a VQC typically involves initializing the qubits in a standard state, $|0\rangle^{\otimes n}$ and then applying a series of quantum gates. These gates are parameterized, meaning their effect on the qubits can be adjusted. Like classical machine learning, the parameters of the circuit are updated by optimizing the cost function. For an input $x$, the output of a VQC is given as

$$f(x;\theta) = \langle 0|U^\dagger(x;\theta)MU(x;\theta)|0\rangle \quad (1)$$

where the unitary matrix $U(x;\theta)$ is a quantum circuit parametrized by $\theta$. $M$ is a Hermitian matrix, also called observable.

For most real-word tasks, such as image recognition, high-performance VQCs are always be hybrid structure composed of classical neural networks and parameterized quantum circuits. In addition to the hybrid structure, data reuploading fashion is another crucial strategy to improve the performance. Indeed, [Schuld *et al*., 2021] has proved that Eq(1) of VQCs with alternate layers of data encoding block and trainable block are mathematically equal to the Fourier series of the input, and with enough data encoding repetition, such models are universal function approximators.

### 2.2 Supernet

In QAS, evaluating performances of all candidate circuits by training is too costly. The primary goal of utilizing Supernet is to efficiently explore the search space by conducting low-cost evaluation. For example, a Supernet can be the strongly entangling circuits with many repeated layers as shown, because it is nature to consider circuits that prepare strongly entangled quantum states are useful to the problem [Schuld *et al*., 2020]. Then, all candidate circuits are restricted to subsets of Supernet. For QuantumNAS [Wang *et al.*, 2022a], the parameters of Supernet are trained by iteratively sampling and updating a subset of parameters. When the training phase is completed, the gate parameters will be inherited to the candidate circuit and then perform evaluation without training.

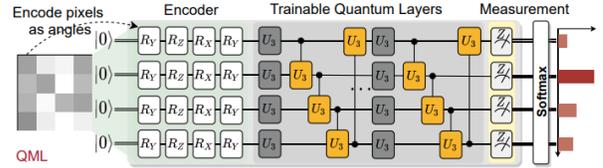

Figure 1. Example SuperCircuit for MNIST. [Wang *et al*., 2022a]

However, one of the primary challenges is the inherent rigidity in its structure. The fixed topology of a Supernet will restrict the search space, omitting a lot of circuit architectures that could offer superior performance in some cases. Considering the MNIST experiment in [Wang *et al*., 2022a], as shown in Figure 1, where Supernet is also called SuperCircuit. If pixels in column 2 and column 3 of the image are exchanged, then QuantumNAS picks the best circuit, which is 8-layer with 62 trainable gates as the final result to handle this exception. However, we found that we could achieve similar or even better performance by modifying the connectivity of qubit 3. The final circuit has only 30 trainable gates, as shown in below.

Obviously, the above result stems from the initial architecture of the Supernet, which constrains the variety of sub-circuits that can be explored, indicating the search for the Supernet of Supernet is inevitable.

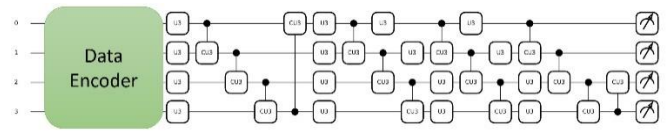

Figure 2. Circuit for the modified MNIST experiment

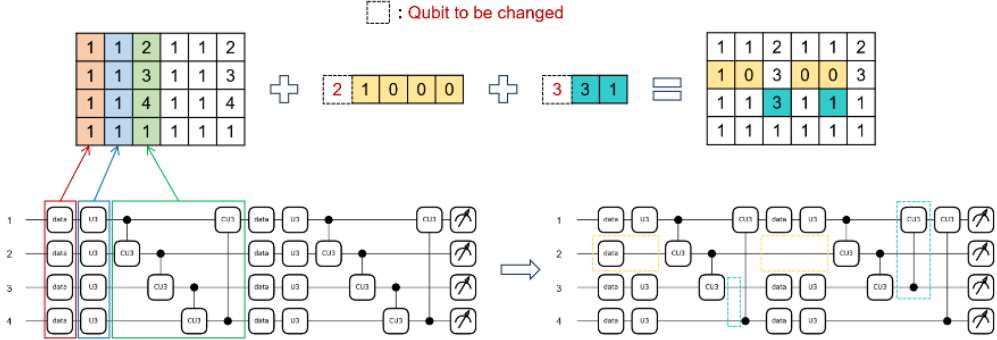

Figure 3. Diagram of circuit structures and encodings

## 2.3 Monte Carlo Tree Search

Monte Carlo Tree Search (MCTS) is a reinforcement learning algorithm that has shown great success in fields of requiring complex decision-making, like strategic games and neural architecture search (NAS) [Silver *et al*., 2016]. In the context of LaNas [Wang *et al*., 2021], each search iteration is composed of two phases, learning phase and search phase.

The goal of the learning phase is to partition the entire search space into several good and bad sets. To this end, a full binary tree is constructed where the root represents the entire search space. At each node of the tree, a linear binary classify is learned from all samples that fall into that node. Then, the whole space $\Omega$ is split recursively into good and bad regions storing in leaves

In the search phase, when a leaf node is selected, whose space are sampled and evaluate them. Keep selecting the leftmost leaf is not a good strategy as classifiers may be not accurate. This problem can be addressed by the UCB score. The basic idea is that if a node has never been visited or infrequently visited, it is necessary to explore the region it relates to.

However, the performance of MCTS heavily depends on the balance between exploration and exploitation, and the quality of evaluation. In complex problems like NAS, finding the right balance and an effective evaluation policy are non-trivial challenges that require careful tuning.

## 3. Qubit-Wise Search method

### 3.1 Overview

The main idea of the QWAS is that in each time step, only a selected part of circuit is modified. This part includes data-uploading gates, single-qubit gates and two-qubit gates acting on a specified qubit only. Unlike the conception of sampling a subset of supernet, QWAS is to reach the better solution by progressively altering the architecture qubit by qubit.

We call the initial circuit as BaseNet. Imagine the BaseNet is a random picture, what QWAS is going to do is make this picture meaningful in limited steps by changing only one row of pixels per step.

### 3.2 Circuit Structures and Encodings

First of all, we describe how to encode the BaseNet like an image. As shown in Figure 3, for an $n$-qubit quantum circuit with $m$ layers, we translate it into a vector of size $n \times m$. Each variable can only take one of a set of predefined discrete integers which represents some structural attribute of the circuit, such as the category and existence of a quantum gate or the connectivity. In this paper, categories of gates are fixed, that is, we choose universal rotation gate $U(3)$ and controlled universal rotation gate $CU(3)$ for single-qubit and two-qubit gates respectively. The kind of gates within data encoding block is also fixed according to the task.

The architecture of the BaseNet can be arbitrary. Generally speaking, it is a good choice to start with the strongly entanglement ansatz shown in the left of Figure *3*. We call this kind of BaseNet as "SuperBase". However, our experiments indicate that starting with a random structure also works well. As quantum circuits are now encoded as a matrix, it is interesting to display them as pixel images.

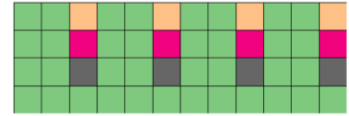

Figure 4. The pixel image of SuperBase

For most real-word tasks, high-performance VQCs are always be hybrid structure made of classical neural networks and parameterized quantum circuits. Due to limits of NISQ, quantum part is far smaller than classical part, if we train the classical part in the search process, the influence of quantum circuit will be ignored. So, the structure and parameters of classical NN are fixed during the search.

### 3.2 Search Strategy

The search strategy splits in two phases: single-qubit gates search and two-qubit gates search. The details are list as follows:
1. Phase one: each sample of the search space is encoded as follows:
$$[q, d@1, s@1, d@2, s@2, \ldots]$$

where $q$ is the position of qubit to be changed. $d@1(s@1)$ is the state of data(single) gates in layer 1, and so on.

2. Phase two: the format of each sample is:
$$[q, t@1, t@2, t@3 ...]$$
where $t@1$ is the position of target qubit in the 1$^{st}$ layer when control qubit is $q$, and so on.

For example, in phase one, the sample [3, 1, 1, 0, 1] means that the data gate on qubit 3 in the 2nd layer is removed as $d@2 = 0$. In phase two, the sample [1, 2, 3] means that in the first layer, qubit 1 is connected to qubit 2 by a CU gate, and connected to qubit 3 by another CU in the second layer. Noticed that if the target number is equal to the qubit number, the corresponding two-qubit gate in that layer is omitted.

Each phase is composed of several stages. In every stage, a batch of samples from search space are acting on the output of the previous stage. The output is one of the best architectures obtained in the current stage, as shown in Algorithm 1.

---
**Algorithm 1** Phase
**Input**: search_space, baseNet
**Parameter**: stages
**Output**: Top N architectures
1: Random pick M samples from search_space
2: **while** iter < stages
3:   Train Monte Carlo Tree in several epochs
5:   Pick the best sample
6:   update BaseNet
9: **end while**
10: **return** Top N architectures

---

After a phase is complete, we switch to another phase to continue this process. Alternating the use of the above two phases can lead the BaseNet to any desired architecture progressively. The key observation is that in each phase, the new architecture is a slight modification of the previous, which requires limited training steps to evaluate its performance. In fact, our experiments show that if the new architecture inherits the parameters of the previous BaseNet, only one epoch of training is enough for all experiments taken. The algorithm of QWAS is shown in algorithm 2.

### 3.3 Classifiers for MCTS

From a high point of view, what we have done is to decompose a large search space into two far smaller disjoint subspaces, $S = S_1 \oplus S_2$, where $S_1$ is the subspace of single-qubit quantum gate configuration and $S_2$ is that of two-qubit. This strategy sometimes will makes it easy for the results to fall into the local optima. So, MCTS is used to balance the exploration and exploitation of the search. The key point is that we build a binary tree shared by two phases, which capture the potential relationships between them.

To construct a successful binary tree of MCTS, the classifiers utilized to partition a node have a significant impact. The linear regressor employed by the original LaNas taking flattened encoding vectors as inputs, which completely ignoring

---
**Algorithm 2** Qubit-Wise Architecture Search
**Input**: search_space_1, search_space_2
**Parameter**: epoch
**Output**: Top N architectures
1: Initial a binary tree with height H
2: **while** iter < epoch **do**
4:   **if** iter % 2 = 0 **then**
5:     baseNet = Phase(search_space_1, baseNet)
6:   **else**
7:     baseNet = Phase(search_space_2, baseNet)
8:   **end if**
9: **end while**
10: **return** baseNet

---

the topology and sequence of quantum interaction among qubits, are proved too simple to learn the mapping between architecture codes and their corresponding performance metric.

Since we have transformed architectures into two-dimensional data according to their spatial and temporal relationship, by employing a *m-by-n* convolutional kernel, we can simultaneously capture at most $mn$ quantum gates enclosed within the kernel. Also, by inputting network into RNN in column-major order, the temporal relationship will be learned. Hence, for a circuit with many layers, a.k.a. an encode matrix with large column numbers, we apply CNN on the input to reduce the number of columns, then use RNN to predict the results, as shown in Figure 4.

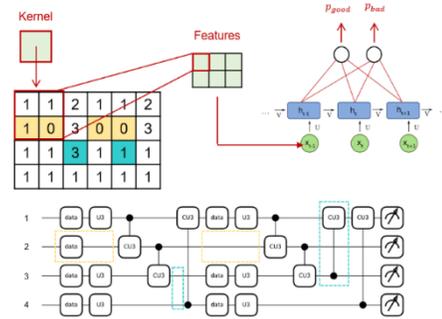

Figure 5. Classifiers for MCTS

### 3.4 Noise-adaptive exploration term

The promise of VQA is can be employed on real noise quantum hardware of today or near future. From the architecture design's point of view, firstly, less quantum gates mean more robustness to the quantum noise. Secondly, less parameterized quantum gates mean more generalization and trainability.

So, our idea is to restrict the number of quantum gates, especially two-qubit gates, within the search process. To this end, the following exploration term $\epsilon$ is introduced into MCTS

$$\epsilon = c_1 N_d + c_2 N_s + c_3 N_e \quad (2)$$

where $N_d, N_s, N_e$ are structure parameters which represent the mean number of data gates, single-qubit gates and two-qubit gates within a node respectively. $c_0, c_1$ and $c_2$ are corresponding coefficients, also called penalty values. With this architecture exploration term, for current node $s$, whose UCB score in the search phase becomes

$$UCB(s,a) = V(s,a) + c_0\sqrt{\frac{2\log N(s)}{N(s,a)}} - \epsilon \quad (3)$$

Where $V(s,a)$ represent the value of child node, $N(s)$ and $N(s,a)$ are the visit numbers of current node and child node respectively.

By adjusting the hyper-parameter $c = (c_0, c_1, c_2, c_3)$ for each node in the MCTS, the dynamics of the search process will be different. Ideally, $c_1, c_2$ and $c_3$ should be optimized by running on real quantum hardware or their noise model to reflect the true performance of a selected circuit.

## 4. Experiments

In this section, we present our experimental results on two real-world tasks, multimodal fusion and image recognition. All the experiments are conducted via the TorchQuantum [Wang *et al.*, 2022b], a fast quantum computing simulation toolkit. The code is available at the paper's repository.

### 4.1 Multimodal Fusion

In our study, we employ the CMU-MOSI dataset to assess our proposed QWAS in the multimodal task of sentiment analysis. The metric evaluated in all experiments is the mean absolute error (MAE).

In this experiment, we construct a tree with a height of 5, culminating in 16 leaf nodes which partition the search space into 16 distinct subregions. We initialize this tree by training 200 randomly chosen architectures. The coefficient $c_0$ in the Upper Confidence Bound (UCB) computation is set to 0.2.

**Single-Pass Fine-Tuning of Entangled Gates (SP)**. In the beginning, we thoroughly train a hybrid fusion model with a SuperBase as described in Section 3.2 with the configuration 7-qubit and 5-layer, and save all its parameters.

To monitor and analyze the search process, after partitioning the search space at the end of each iteration, we randomly select 100 architectures from the first 4, the median and the last 4 leaf nodes individually, train them and record their average performance metric. Such an analysis can provide insights into the performance variance across the partitioned subregions. As depicted in Figure 7, there is a notable trend in the average MAE across the leaf nodes, progressing from left to right, after 40 iterations of training. Specifically, on average, the models situated in the first four leaf nodes exhibit significantly superior performance in comparison to those in the last four nodes, while the models in the median leaf node register an average MAE of 1.321 which is close to that of randomly selected samples, as listed in Table 1. This empirical evidence demonstrates the effectiveness of our QWAS algorithm in partitioning the search space based on model performance.

Then, we compare the sampling efficacy of the QWAS algorithm with random sampling by analyzing the sample distribution over 40 iterations across different MAE intervals, as shown in Figure 6, where the sample count from QWAS is normalized to match the base of 200 random samples. Observing the results, it is evident that for lower MAE intervals (ranging from 1.19 to 1.29), QWAS samples exhibit a notably advantage in quantity compared to the random counterparts. To evaluate our algorithm's performance more quantitatively against random sampling, we calculate the average MAE of the final 200 samples across 40 iterations. This data is then compared to the results from random sampling. Our findings reveal that the QWAS algorithm consistently outperforms random sampling, achieving a significantly lower average MAE of 1.269, compared to 1.329 for the random samples. These results suggest a superior efficiency of QWAS in sampling high-performing models.

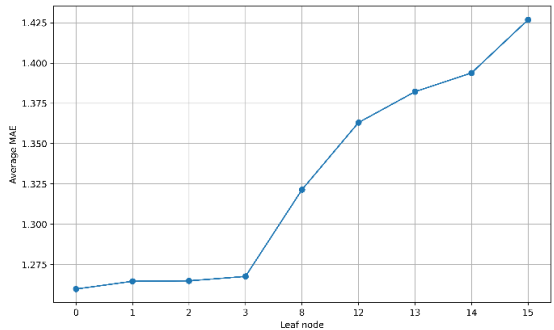

Figure 7. Average MAE of the leaf node

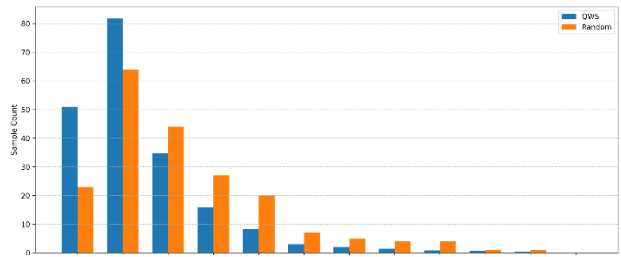

Figure 6. The sample distribution within different MAE intervals

**Multiple-Pass Fine-Tuning of Entangled Gates (MP)**. The advantage of our QWAS algorithm is the scalability to multiple architectural fine-tuning. We report results of 3 subsequent stages. In each subsequent stages, we take the best model discovered at the previous stage as our new base model. So, 3 qubits can be changed at most. Each stage takes 10 iterations.

We quantify the distribution of sample counts across various MAE intervals at each stage, with the results being illustrated in Figure 9. The sample count at each stage is normalized to ensure a fair comparison. The figure demonstrates a clear improvement in the algorithm's capacity to identify high-performing models with each additional fine-tuning iteration, as evidenced by the growing number of samples in the lowest MAE interval (ranging from 1.18 to 1.23). Additionally, we calculate the average MAE of the final 200 samples at each stage. The results are listed in Table 1. The table clearly indicates that as the search stage progresses, there is a discernible improvement in the average performance of the collected samples. This enhancement is quantitatively reflected by a decrease in MAE, which steadily declines from 1.291 at stage 1 to 1.263 at stage 3. This finding indicates that progressively refining entangled gates can yield steady enhancements in the model's performance.

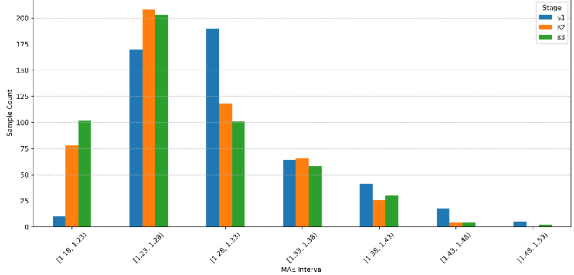

Figure 9. The sample distribution within different MAE intervals

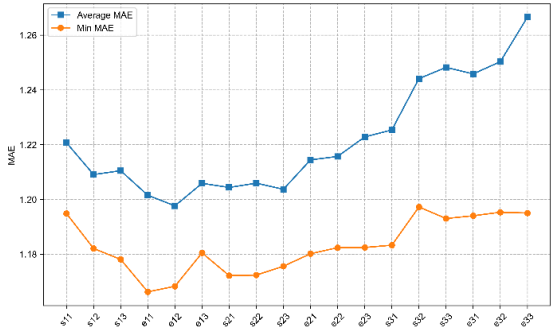

Figure 8. Average and minimum MAE in each stage

**Interleaved of Phase one and Phase two**. Furthermore, we introduce modifications to the single-qubit gates for both data uploading and training. In each iteration, the single-qubit gates are adjusted three times, each time on a different qubit, followed by a similar fine-tuning process for the entangled gates. Subsequent to each tuning phase, one of the top 2 models, as identified in the current search, is randomly selected for the ensuing fine-tuning step. This iterative procedure is executed a total of three times.

We calculate the average and minimum MAE of the models identified in each stage, and the results are depicted in Figure 8, where $s_{ij}$ and $e_{ij}$ respectively denote the $j$th adjustment in the $i$th iteration of the single-qubit gates and the entangled gates. We can see that the average MAE exhibits a pattern of initial decline followed by an increase across the stages. In only one iteration, the best model discovered outperform those of all preceding experiments, achieving an MAE of 1.166, as detailed in Table 1.

### 4.2 Image Recognition

We conduct experiments on 2 tasks of image recognition, MNIST-4 and Fashion-4. The information is consistent with those used in QuantumNAS [Wang et al., 2022]. On both tasks, 95% images in 'train' split are used as the training set and 5% as the validation set. The $28 \times 28$ input images are center-cropped to $24 \times 24$ and down-sampled to $4 \times 4$ with average pooling to adapt to the VQC input size. The BaseNet we used in the following experiments are 4-qubit VQCs with a fixed layer number of 4. The building blocks are data encoders, U3 and CU3 gates.

**Comparison with QuantumNAS**. We first report our results compared with the QuantumNAS. The main difference is that QuantumNAS uses a SuperCircuit up to 8 layers at most to search for subcircuits. The results are shown in Table 2. After 51 iterations and evaluating 700 samples, Our QWAS algorithm outputs best circuits achieve accuracy of 84.51% on MNIST-4 and 84.50% on Fashion-4 starting from a random circuit and the SuperBase, respectively. These results are superior to the accuracy of 82.27% on MNIST-4 and 81.45% on Fashion-4 achieved by QuantumNAS. Additionally, we calculate the number of single-qubit gates and the entangled gates required to achieve the best accuracy in both QWAS and QuantumNAS, where the number of data gates is converted into the equivalent number of U3 gates at ration 1:1.33. As listed in Table 2, our discovered circuits require even less gate count in both single-qubit gate and entangled gate than that in QuantumNAS, which brings a 33.3% and 36.6% reduction in parameters, respectively. This is beneficial for the low noise implementation of the circuits on practical quantum devices and the reduction in training time of the models.

| method | settings | MAE |
|---|---|---|
| SP | random | 1.329 |
| | QWS-200 | 1.269 |
| | QWS-best | 1.194 |
| MP | QWS-s1 | 1.291 |
| | QWS-s2 | 1.276 |
| | QWS-s3 | 1.263 |
| | QWS-best | 1.185 |
| IL | QWS-best | 1.166 |

Table 1: Results of QWS for multimodal fusion task

| Method | Task | ACC | #single | #enta | #param |
|---|---|---|---|---|---|
| QuantumNAS | MNIST | 82.27% | 25.33 | 19 | 117 |
| | Fashion | 81.45% | 26.33 | 20 | 123 |
| QWAS | MNIST | 84.51% | 21.67 | 15 | 78 |
| | Fashion | 84.50% | 25 | 13 | 78 |

Table 2: Compared with QuantumNAS

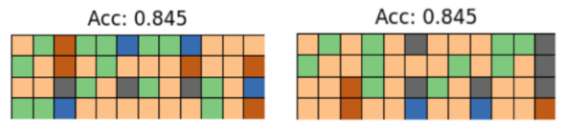

Figure 10. Pixel images of best circuits. L: MNIST. R: FASHION

**Impact of the starting circuit.** As aforementioned, our QWAS can search for an architecture from any starting circuit even an empty circuit. However, we empirically observe that initiating the search with an empty circuit will result in prolonged periods of low performance. Thus, we experiment with the SuperBase and a random circuit. We sequentially conduct three iterations of both Phase one and Phase two, with each phase comprising 9 iterations, resulting in a total of 54 iterations. Figure 5 illustrates the highest model accuracy achieved in each iteration of the search. It is evident that starting with the SuperBase is a judicious choice, as fine-tuning on a strongly entangled structure rapidly leads to the identification of high-performance models. However, experimental results indicate that even when beginning with a lower-performance random circuit, we can swiftly adjust it to achieve a high-performance model, potentially surpassing searches that commence with the SuperBase, as can be seen in the results presented in Table 2.

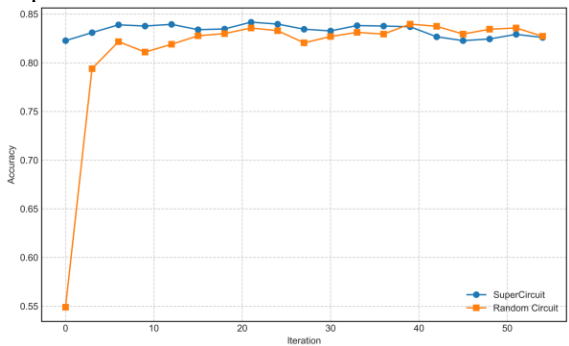

*Figure 11 the highest model accuracy achieved in the search*

**Impact of the exploration term and MCTS.** Direct searches often lead to local optima and typically result in circuits with a higher number of gates. The introduction of an exploration term aims to mitigate this issue. Experiments incorporating the exploration term are conducted with searches starting from both the SuperBase and a random circuit, and the results are presented in Table 3. The three numbers associated with the exploration term quantify the penalty levels for data gates, single-qubit gates, and entangled gates, respectively. The experimental findings suggest that in many cases, the introduction of a penalty can enhance the performance of the models identified through the search. On the MNIST-4 task, searches that begin with a random circuit even see an accuracy improvement of 4.36%. Although a higher penalty does not always lead to regions of better samples but does indeed tend to steer the search towards regions containing circuits with fewer quantum gates.

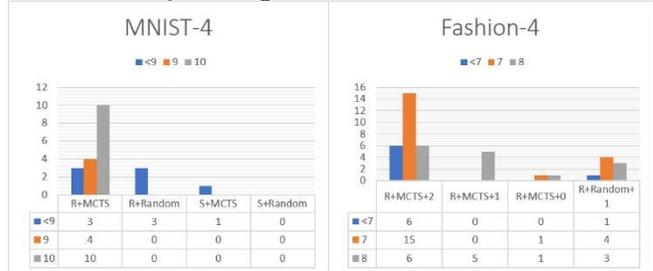

Figure 12. The number of samples in various two-qubit gate condition

As listed in Figure 12, we count the number of circuits, out of the top 100 sampled in searches with varying penalty levels, that have fewer than 7, 7 and 8 entangled gates respectively. "R+MCTS+0" means random BaseNet, using MCTS and 0 penalty. Here, 0 means no exploration term, 1 and 2 are two level of exploration in the ascending order. Obviously, MCTS with large punish term will lead to find more smaller circuits. Moreover, we compare the effect of MCTS and random search under the QWAS strategy. "R+Random+1" represent random search on random circuit, at the penalty level 1. In fact, as the proposed strategy performs well in these tasks, which also makes the random search competitive. However, Figure 12 shows when penalty increases, MCTS is still better than random search to find more compact structure. This is also true in MNIST, comparing with others, "R+MCTS" is best at the same penalty level.

## 5. Conclusions and Outlooks

In this paper, we propose a novel qubit-wise architecture search (QWAS) method, which progressively optimizes one-qubit configuration per stage, and combine with MCTS to partition the search space of next stages into several good and bad subregions. The numerical experimental results show that our proposed method can balance the exploration and exploitation of circuit performance and size. As far as we know, in the context of quantum machine learning, our method has achieved SOTA in MNIST, Fashion and MOSI in terms of accuracy and circuit size. More importantly, since our implementation can start with any circuit architectures and improve their performance, which has wider applications.

For future directions, we plan to apply QWAS to larger-scale applications and search for quantum circuits with more than 10 qubits. We believe with the increase in scale, this method will become more superior. Moreover, as quantum circuits can be encoded as an image, this suggests us to extend our search method from row-by-row to area-by-area fashion. MCTS or other reinforcement learning method can be utilized to sample a candidate area based on the current state. Finally, how to integrate Supernet-based methods with QWAS to develop a training-less search method is a promising direction.

| Task | BaseNet | Exploration Term | ACC |
|---|---|---|---|
| MNIST-4 | Super | No | 83.91% |
| | Super | [0.001, 0.002, 0.003] | 83.71% |
| | Random | No | 80.15% |
| | Random | [0.001, 0.002, 0.003] | **84.51%** |
| Fashion-4 | Super | No | 84.18% |
| | Super | [0.001, 0.002, 0.003] | **84.50%** |
| | Random | No | 83.98% |
| | Random | [0.001, 0.002, 0.003] | 84.05% |

Table 3. Results with different BaseNet and exploration terms

# Acknowledgments
.

# References

[Preskill, 2018] Preskill, John. "Quantum computing in the NISQ era and beyond." *Quantum* 2 (2018): 79.

[Mitarai *et al*., 2018] Mitarai, Kosuke, et al. "Quantum circuit learning." *Physical Review A* 98.3 (2018): 032309.

[Kandala *et al*., 2017] Kandala, Abhinav, Antonio Mezzacapo, Kristan Temme, Maika Takita, Markus Brink, Jerry M. Chow, and Jay M. Gambetta. "Hardware-efficient variational quantum eigensolver for small molecules and quantum magnets." *nature* 549, no. 7671 (2017): 242-246.

[Cao *et al*., 2019] Cao, Yudong, Jonathan Romero, Jonathan P. Olson, Matthias Degroote, Peter D. Johnson, Mária Kieferová, Ian D. Kivlichan et al. "Quantum chemistry in the age of quantum computing." *Chemical reviews* 119, no. 19 (2019): 10856-10915.

[Cong *et al*., 2019] Cong, Iris, Soonwon Choi, and Mikhail D. Lukin. "Quantum convolutional neural networks." *Nature Physics* 15, no. 12 (2019): 1273-1278.

[Henderson *et al*., 2020] Henderson, Maxwell, Samriddhi Shakya, Shashindra Pradhan, and Tristan Cook. "Quanvolutional neural networks: powering image recognition with quantum circuits." *Quantum Machine Intelligence* 2, no. 1 (2020): 2.

[Schuld *et al*., 2020] Schuld, Maria, Alex Bocharov, Krysta M. Svore, and Nathan Wiebe. "Circuit-centric quantum classifiers." *Physical Review A* 101, no. 3 (2020): 032308.

[Pirhooshyaran *et al*., 2021] Pirhooshyaran, Mohammad, and Tamas Terlaky. "Quantum circuit design search." *Quantum Machine Intelligence* 3 (2021): 1-14.

[Holmes *et al*., 2022] Holmes, Zoë, Kunal Sharma, Marco Cerezo, and Patrick J. Coles. "Connecting ansatz expressibility to gradient magnitudes and barren plateaus." *PRX Quantum* 3, no. 1 (2022): 010313.

[Elsken *et al*., 2019] Elsken, Thomas, Jan Hendrik Metzen, and Frank Hutter. "Neural architecture search: A survey." *The Journal of Machine Learning Research* 20, no. 1 (2019): 1997-2017.

[Silver *et al*., 2016] Silver, David, Aja Huang, Chris J. Maddison, Arthur Guez, Laurent Sifre, George Van Den Driessche, Julian Schrittwieser et al. "Mastering the game of Go with deep neural networks and tree search." *nature* 529, no. 7587 (2016): 484-489.

[Wang *et al*., 2020] Wang, Linnan, Rodrigo Fonseca, and Yuandong Tian. "Learning search space partition for black-box optimization using monte carlo tree search." *Advances in Neural Information Processing Systems* 33 (2020): 19511-19522.

[Wang *et al*., 2021] Wang, Linnan, Saining Xie, Teng Li, Rodrigo Fonseca, and Yuandong Tian. "Sample-efficient neural architecture search by learning action space." arXiv preprint arXiv:1906.06832 (2019). *IEEE Transactions on Pattern Analysis and Machine Intelligence* 44(9):5503–5515.

[Du *et al*., 2022] Du, Yuxuan, Tao Huang, Shan You, Min-Hsiu Hsieh, and Dacheng Tao. "Quantum circuit architecture search for variational quantum algorithms." *npj Quantum Information* 8, no. 1 (2022): 62. [Wang *et al*., 2022a] Wang, Hanrui, Yongshan Ding, Jiaqi Gu, Yujun Lin, David Z. Pan, Frederic T. Chong, and Song Han. "Quantumnas: Noise-adaptive search for robust quantum circuits." In *2022 IEEE International Symposium on High-Performance Computer Architecture (HPCA)*, pp. 692-708. IEEE, 2022. [Alexeev *et al*., 2020] Alexeev, Yuri, Sami Khairy, Ruslan Shaydulin, Lukasz Cincio, and Prasanna Balaprakash. "Reinforcement learning for finding qaoa parameters." *Bulletin of the American Physical Society* 65 (2020).

[Zhang *et al*., 2020] Zhang, Shi-Xin, Chang-Yu Hsieh, Shengyu Zhang, and Hong Yao. "Differentiable quantum architecture search." *Quantum Science and Technology* 7, no. 4 (2022): 045023.

[Zoph and Le, 2016] Zoph, Barret, and Quoc V. Le. "Neural architecture search with reinforcement learning." *arXiv preprint arXiv:1611.01578* (2016).

[Liu *et al*., 2018] Liu, Hanxiao, Karen Simonyan, and Yiming Yang. "Darts: Differentiable architecture search." *arXiv preprint arXiv:1806.09055* (2018).

[Rosenhahn *et al*., 2023] Rosenhahn, Bodo, and Tobias J. Osborne. "Monte Carlo graph search for quantum circuit optimization." *Physical Review A* 108.6 (2023): 062615

[Wang *et al*., 2023] Wang, Peiyong, et al. "Automated quantum circuit design with nested monte carlo tree search." *IEEE Transactions on Quantum Engineering* (2023).

[Zhou *et al*., 2022] Zhou, Xiangzhen, Yuan Feng, and Sanjiang Li. "Quantum circuit transformation: A Monte Carlo tree search framework." *ACM Transactions on Design Automation of Electronic Systems (TODAES)* 27.6 (2022): 1-27

[Meng *et al*., 2021] Meng, Fan-Xu, et al. "Quantum circuit architecture optimization for variational quantum eigensolver via monto carlo tree search." *IEEE Transactions on Quantum Engineering* 2 (2021): 1-10.

[Ostaszewski *et al*., 2021] Ostaszewski, Mateusz, Lea M. Trenkwalder, Wojciech Masarczyk, Eleanor Scerri, and Vedran Dunjko. "Reinforcement learning for optimization of variational quantum circuit architectures." *Advances in Neural Information Processing Systems* 34 (2021): 18182-18194.

[Wang *et al*., 2022b] Wang, Hanrui, Zhiding Liang, Jiaqi Gu, Zirui Li, Yongshan Ding, Weiwen Jiang, Yiyu Shi, David Z. Pan, Frederic T. Chong, and Song Han. "TorchQuan-tum Case Study for Robust Quantum Circuits." In *Proceedings of the 41st IEEE/ACM International Conference on Computer-Aided Design*, pp. 1-9. 2022.